\documentclass[reprint,amsmath,amssymb,aps,prb,longbibliography]{revtex4-2}

\usepackage{graphicx}
\usepackage{amsmath}
\usepackage{xcolor}
\usepackage{dcolumn}
\usepackage{enumitem}
\usepackage{nicefrac}

\begin{document}

\title{Magnetization dynamics in proximity-coupled superconductor-ferromagnet-superconductor multilayers. Part II.}

\author{I.~A.~Golovchanskiy$^{1,2,3,*}$, N.~N.~Abramov$^{2}$, V.~V.~Ryazanov$^{1,2,4}$, A.~A.~Golubov$^{5}$, V.~S.~Stolyarov$^{1,2,3}$}

\affiliation{
$^1$ Moscow Institute of Physics and Technology, State University, 9 Institutskiy per., Dolgoprudny, Moscow Region, 141700, Russia; 
$^2$ National University of Science and Technology MISIS, 4 Leninsky prosp., Moscow, 119049, Russia; 
$^3$ Dukhov Research Institute of Automatics (VNIIA), 127055 Moscow, Russia; 
$^4$ Institute of Solid State Physics (ISSP RAS), Chernogolovka, 142432, Moscow region, Russia; 
$^5$ Faculty of Science and Technology and MESA+ Institute for Nanotechnology, University of Twente, 7500 AE Enschede, The Netherlands.
}%

\begin{abstract}
In this work, we the study magnetization dynamics in superconductor-ferromagnet-superconductor thin-film structures .
Results of the broad-band ferromagnetic resonance spectroscopy are reported for a large set of samples with varied thickness of both superconducting and ferromagnetic layers in a wide frequency, field, and temperature ranges.
Experimentally the one-dimensional anisotropic action of superconducting torque on magnetization dynamics is established; its dependence on thickness of layers is revealed.
It is demonstrated that experimental findings support the recently-proposed mechanism of the superconducting torque formation via the interplay between the superconducting kinetic inductance and magnetization precession at superconductor-ferromagnet interfaces.
\end{abstract}

\maketitle

\section{Introduction}

Advantages from hybridization of antagonistic superconducting and ferromagnetic orders in electronics and spintronics have been repeatedly demonstrated in past decades \cite{Linder_NatPhys_11_307}. 
The interplay between ferromagnetic and superconducting spin orders enables manipulation with spin states and leads to a development of various electronic and spintronic elements, including  superconductor-ferromagnet-superconductor (S-F-S) Josephson junctions \cite{Linder_NatPhys_11_307,Ryazanov_PRL_86_2427,Weides_APL_89_122511}, superconducting phase shifters \cite{Yamashita_SRep_10_13687,Feofanov_NatPhys_6_593}, memory elements \cite{Golovchanskiy_PRB_94_214514,Vernik_IEEETAS_23_1701208,Karelina_JAP_130_173901}, F-S-F–based spin valves \cite{Gingrich_NatPhys_12_564,Lenk_PRB_96_184521}, Josephson diodes \cite{Jeon_NatMat_2022} and more complex long-range spin-triplet superconducting systems \cite{Robinson_Sci_329_59,Banerjee_NatComm_5_4771,Wang_NatPhys_6_389,Glick_SciAdv_4_eaat9457,Jeon_NatMat_20_1358}.

Recently application capabilities of S-F hybridization have been expanded by demonstrations of its prospects in magnonics. 
Magnonics is a growing field of research which offers approaches for the transfer and processing of information via spin waves.
A good overview of various potential applications and recent advances in magnonics can be found in Refs.~\cite{Chumak_IEEE_58_0800172,Pirro_NatRevMat_6_1114,Barman_JPCM_33_413001,Chumak_NatPhys_11_453,Lenk_PhysRep_507_107,Csaba_PLA_381_1471,Serga_JPDAP_43_264002,Kruglyak_JPDAP_43_264001} and references therein.
In development of magnonic systems one of principle requirements is engineering of appropriate spin-wave dispersion.  

Various wide-range manipulations with the spin-wave dispersion have been demonstrated at cryogenic temperatures when coupling magnonic systems with superconductors.
For instance, interaction of a magnonic media with the superconducting vortex matter allows to 
form and tune forbidden bands at sub-micrometer wavelength which matches the parameter of the vortex lattice \cite{Dobrovolskiy_NatPhys_15_477}, or to excite exchange spin waves via driving the vortex lattice with the electric current \cite{Dobrovolskiy_arXiv_2103.10156}. 
Also, magnetostatic interaction of spin waves with superconducting Meissner currents in thin-film hybrid structures modifies substantially the spin-wave dispersion \cite{Golovchanskiy_AFM_28_1802375,Golovchanskiy_JAP_124_233903} and can be used to form magnonic crystals \cite{Golovchanskiy_AdvSci_6_1900435} or to gate the magnon current \cite{Yu_2201_09532}.
Remarkably, low speed of electromagnetic propagation in superconductor-insulator-superconductor thin-film structures facilitates achievement of the ultra-strong photon-to-magnon coupling in on-chip hybrid devices \cite{Golovchanskiy_SciAdv_7_eabe8638,Golovchanskiy_PRAppl_16_034029}. 
 

A striking phenomenon in S-F hybrid structures was reported recently in Ref.~\cite{Li_ChPL_35_077401} and studied in more details in Ref.~\cite{Golovchanskiy_PRAppl_14_024086}. 
In superconductor-ferromagnet-superconductor trilayer thin-film structures a radical increase in the ferromagnetic resonance (FMR) frequency occurs in the presence of electronic interaction between superconducting and ferromagnetic layers.
The phenomenon is strong: in some S-F-S structures the highest natural FMR frequencies are reached among in-plane magnetized ferromagnetic systems \cite{Golovchanskiy_PRAppl_14_024086}.
Intriguingly, so far, the mechanism behind the phenomenon remains unestablished.   

In this work, we report a comprehensive experimental study of the phenomenon. 
We report results of FMR spectroscopy for a large set of samples with varied thickness of both superconducting and ferromagnetic layers in a wide frequency, field, and temperature ranges.
We establish an anisotropic one-dimensional action of hybridization-induced torque acting on magnetization dynamics and the dependence of this superconducting torque on the magnetic field.
Experimental results support the recently proposed model by M. Silaev \cite{Silaev}, which explains the phenomenon in S-F-S structures as the outcome of the coupling between magnetization dynamics and superconducting kinetic inductance at S-F interfaces.

\section{Experimental details}

%
\begin{figure}[!ht]
\begin{center}
\includegraphics[width=1\columnwidth]{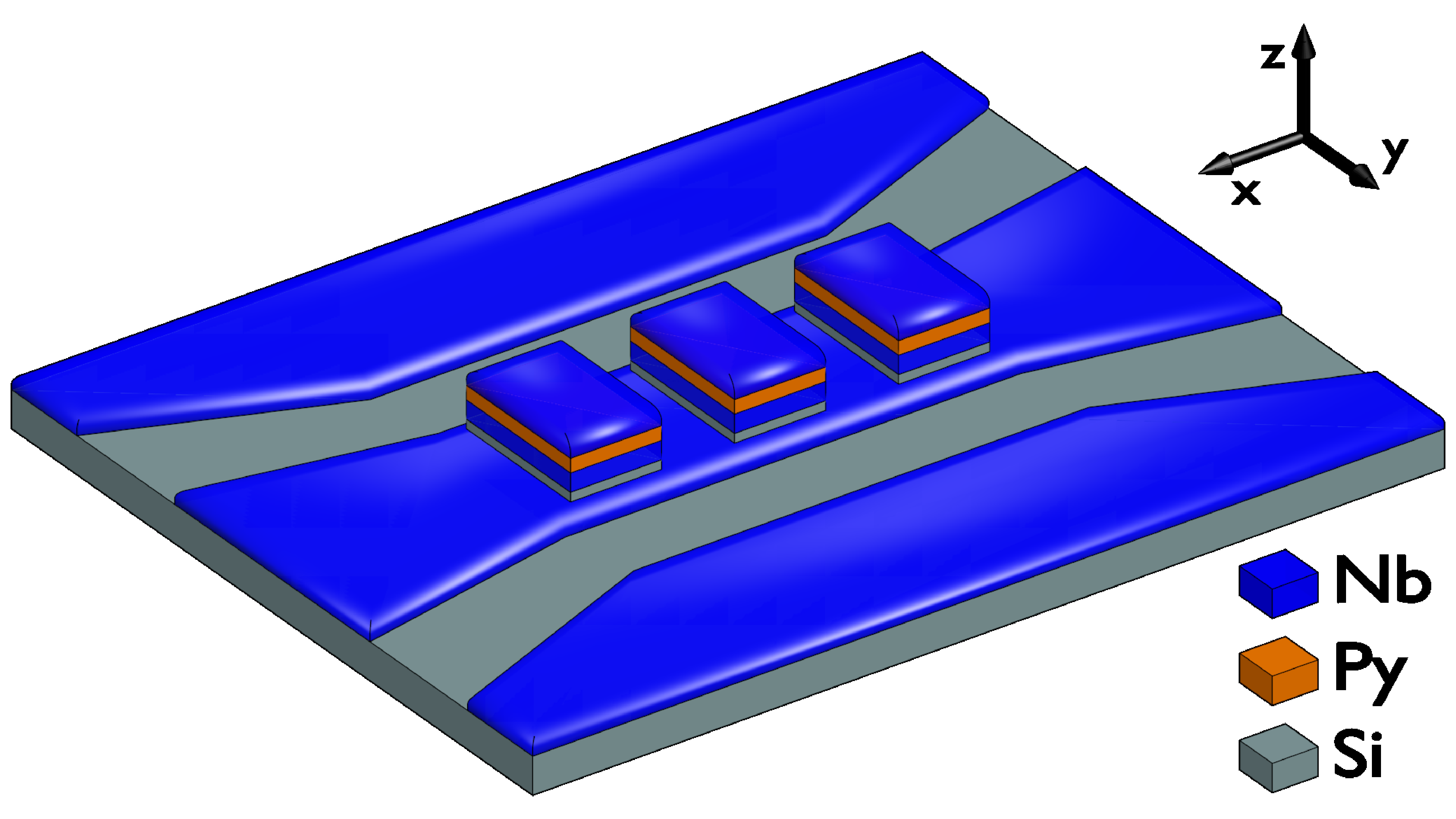}
\caption{
Schematic illustration of the investigated chip-sample (adopted from Ref.~\cite{Golovchanskiy_PRAppl_14_024086}).  
A series of S-F-S film rectangles is placed directly on top of the central transmission line of niobium co-planar waveguide.
Magnetic field $H$ is applied in-plane along the $x$-axis.
}
\label{sam}
\end{center}
\end{figure}

\begin{table*}[!ht]
\begin{center}
\begin{tabular}{|c|c|c|c|c|c|c|c|c|c|}
\hline
Sample ID  & $d_s$ (Nb), nm & $d_F$ (Py), nm & $d_s$ (Nb), nm & $\mu_0 H_a$, mT & $\mu_0 M_{eff}$, T & $\mu_0 H_{s0}$, mT & $T_c$, K & $p$ \\
\hline
S1 & 110 & 10 & 110 & 1.5 & 1.051 & 31.8 & 8.74 & 3.79 \\
\hline
S2 & 120 & 11 & 120 & -3.3 & 1.029 & 39.1 & 8.56 & 3.70 \\
\hline
S3 & 110 & 19 & 110 & -0.15 & 1.100 & 78.8 & 8.90 & 3.48 \\
\hline
S4 & 100 & 35 & 100 & -0.2 & 1.132 & 163.6 & 8.92 & 3.80 \\
\hline
S5 & 140 & 45 & 174 & -0.4 & 1.18 & 193.7 & 9.08 & 4.90 \\
\hline
S6 & 110 & 120 & 110 & -0.1 & 1.123 & 352.6 & 8.82 & 4.56 \\
\hline
S7 & 110 & 350 & 110 & 1.6 & 1.076 & 617.2 & 8.74 & 8.89 \\
\hline
S8 & 41 & 24 & 41 & -0.2 & 1.131 & 37.8 & 7.51 & 3.07 \\
\hline
S9 & 200 & 30 & 50 & -1 & 1.146 & 96.8 & 8.74 & 4.38 \\
\hline
\end{tabular}
\caption{Parameters of studied samples.
The $d_s$ and $d_F$ denote thickness of superconducting and ferromagnetic layers, respectively.
The $H_a$ and $M_{eff}$ correspond to parameters obtained at $T>T_c$ with Eq.~\ref{Kit}.
The $H_{s0}$, $T_c$, and $p$ are obtained by fitting $H_s(T)$ in Fig.~\ref{Fit1} with Eq.~\ref{Hs}.
}
\label{Tab}
\end{center}
\end{table*}

Magnetization dynamics in S-F-S structures is studied by measuring the ferromagnetic resonance absorption spectrum using the VNA-FMR approach \cite{Neudecker_JMMM_307_148,Kalarickal_JAP_99_093909,Chen_JAP_101_09C104,Golovchanskiy_JAP_120_163902}.
A schematic illustration of investigated samples is shown in Fig.~\ref{sam}.
The chip consists of a thin-film superconducting niobium (Nb) co-planar waveguide with 50~Ohm impedance and 82-150-82~$\mu$m center-gap-center dimensions 
and a series of niobium-permalloy(Py=Fe$_{20}$Ni$_{80}$)-niobium (Nb-Py-Nb) film structures with lateral dimensions $X\times Y=50\times140$~$\mu$m and spacing of 25~$\mu$m along the $x-$axis that are placed directly on top of the central transmission line of the waveguide.
Deposition of Nb-Py-Nb trilayers is performed in a single vacuum cycle ensuring the electron transparency at Nb-Py interfaces. 
Thin Si or AlO$_x$ spacing layer is deposited between Nb co-planar waveguide and the trilayers in order to ensure electrical insulation of the studied samples from the waveguide.
As a result, a series of samples has been fabricated and measured with different thickness of superconducting (S) and ferromagnetic (F) layers (see Tab.~\ref{Tab}).

Microwave spectroscopy of samples was performed by measuring the transmission characteristics $S_{21}(f,H)$ in the closed-cycle cryostat Oxford Instruments Triton (base temperature 1.2 K) equipped with the home-made superconducting solenoid.
Spectroscopy was performed in the field range from -0.22 T to 0.22 T, in the frequency range from 0 up to 20 GHz, and in the temperature range from 2 to 11 K. 
Magnetic field was applied in-plane along the direction of the waveguide in Fig.~\ref{sam}.
FMR spectra at different temperatures were analysed by fitting $S_{21}(f)$ characteristics at specified $H$ and $T$ with the Lorentz curve and, thus, obtaining the dependencies of the resonance frequency on magnetic field $f_r(H)$.


\section{Experimental results}

\begin{figure}[!ht]
\begin{center}
\includegraphics[width=1\columnwidth]{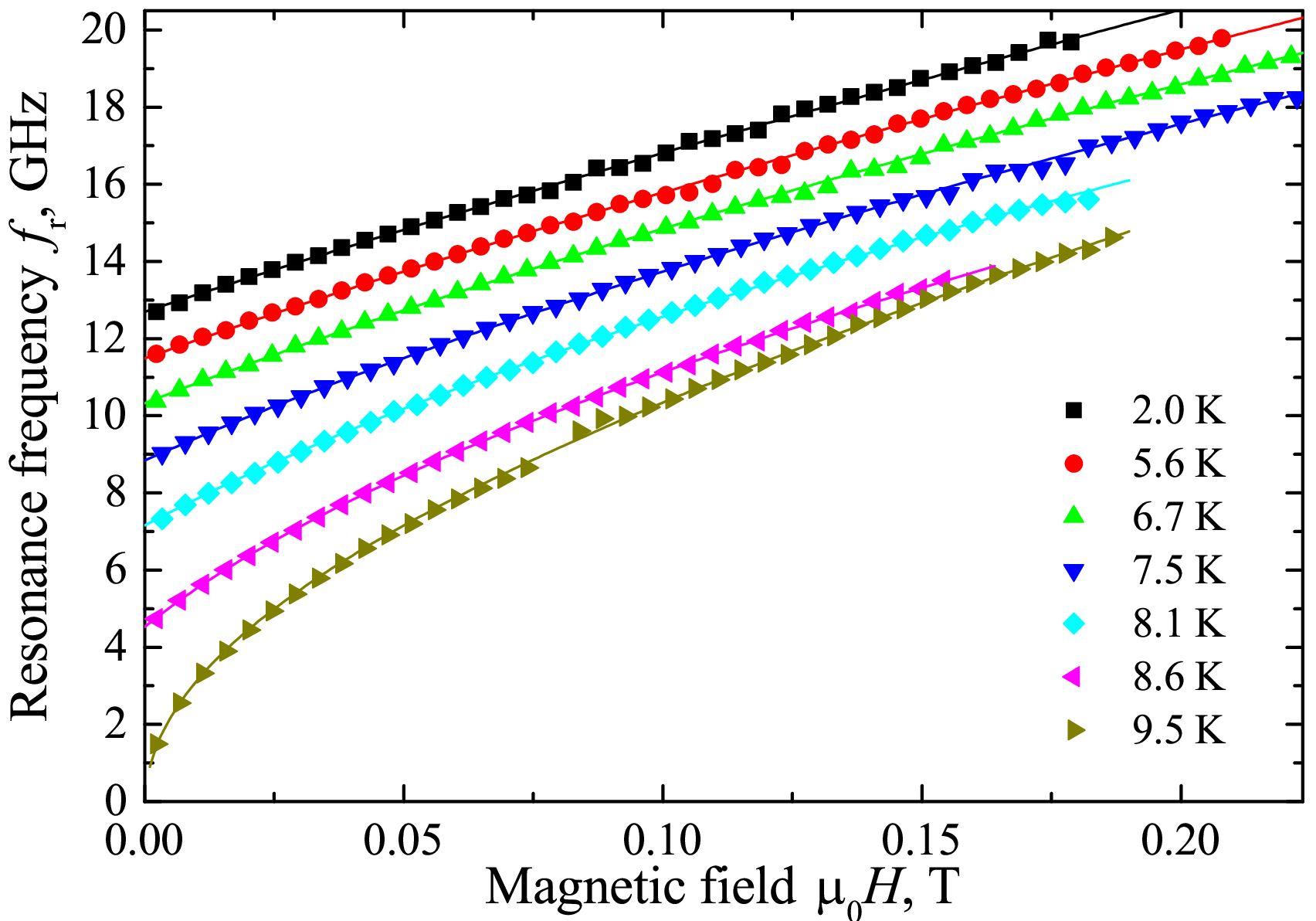}
\caption{
Symbols show experimental dependencies of the FMR frequency on magnetic field $f_r(H)$ for the S5 sample at different temperatures.
Solid lines show corresponding fit of $f_r(H)$ dependencies with Eq.~\ref{Kit} at $T>T_c$ and with Eq.~\ref{Kit_m} at $T<T_c$.
}
\label{Exp1}
\end{center}
\end{figure}

Figure~\ref{Exp1} demonstrates the essence of the studied phenomenon:
it shows dependencies of the FMR frequency on magnetic field $f_r(H)$ for the S4 sample (see Tab.~\ref{Tab}) which is used as the representative example.
The S5 sample consists of 100~nm thick Nb layers and 35~nm thick Py layer. 
Upon decreasing the temperature below the critical temperature of Nb $T_c\approx9$~K the resonance curve $f_r(H)$ shifts gradually to higher frequencies.
For instance, upon decreasing the temperature the frequency of the natural FMR $f_r(H=0)$ increases from about 0.5~GHz at $T\geq 9$~K to about 13~GHz at $T=2$~K.

At $T>T_c$ FMR curves $f_r(H)$ in Fig.~\ref{Exp1} follow the typical Kittel dependence for thin in-plane-magnetized ferromagnetic films at in-plane magnetic field: 
\begin{equation}
\left(2\pi f_r/\mu_0\gamma\right)^2=\left(H+H_a\right)\left(H+H_a+M_{eff}\right)
\label{Kit}
\end{equation}
where $\mu_0$ is the vacuum permeability, $\gamma=1.856\times10^{11}$~Hz/T is the gyromagnetic ratio for permalloy, $H_a$ is the uniaxial anisotropy field that is aligned with the external field, and $M_{eff}$ is the effective magnetization, which includes the saturation magnetization $M_s$ and the out-of-plane anisotropy field.
For all studied samples the fit of FMR curves at $T>T_c$ with Eq.~\ref{Kit} yields negligible anisotropy field $H_a$, the effective magnetization $M_{eff}\approx1.0-1.2$~T, which is close to typical values of the saturation magnetization of permalloy $\mu_0 M_s\approx1$~T, and no noticeable dependence of $H_a$ and $M_{eff}$ on temperature.
Magnetic parameters $H_a$ and $M_{eff}$ for all studied samples are provided in Tab.~\ref{Tab}.
The fit of $f_r(H)$ curves for the S5 sample with Eq.~\ref{Kit} at temperature $T=9.5$~K$>T_c$ is show in Fig.~\ref{Exp1} with yellow solid line.

At $T<T_c$ FMR curves obey a different expression. 
In Refs.~\cite{Golovchanskiy_PRAppl_14_024086, Golovchanskiy_SciAdv_7_eabe8638} it was shown that technically by fitting $f_r(H)$ at $T\ll T_c$ with Eq.~\ref{Kit} the action of superconductivity result in equal but different in sign uniaxial anisotropy field $H_a$ and changes of the effective magnetization $\Delta M_{eff}$:  $H_a =-\Delta M_{eff}$. 
Following the basics of derivation of the Kittel formula \cite{Stancil}, this equality indicates that superconductivity acts on magnetization as the one-dimensional restoring torque along the $y$ axis in Fig.~\ref{sam}, and the fitting function should take the form $f_r^2\sim (H+H_s)(H+M_{eff})$, where $H_s$ is the field of the superconducting torque.
A satisfactory fit with such expression can be obtained for all samples in Tab~\ref{Tab} at temperatures $T\ll T_c$, while for the sample S8, which is based on thin superconducting layers with $d_s<\lambda_L$, this expression is valid in the entire temperature range.
Here $\lambda_L\approx 80$~nm \cite{Golovchanskiy_SciAdv_7_eabe8638,Gubin_PRB_72_064503} is the London penetration depth in bulk niobium at zero temperature

However, as was also shown in Refs.~\cite{Golovchanskiy_PRAppl_14_024086, Golovchanskiy_SciAdv_7_eabe8638} at temperatures $T\lesssim T_c$ the superconductivity-induced uniaxial anisotropy field $H_a$ in Eq.~\ref{Kit} no longer correspond to $-\Delta M_{eff}$.
As it appears, this discrepancy is a rather artificial effect and can be explained by suppression of the superconducting torque upon increasing the magnetic field. 
At higher fields suppression of the field $H_s$ result in reduction of the derivative from the $f_r(H)$ curve.
The fit of such resonance line with the conventional Kittel formula Eq.~\ref{Kit} result in larger reduction of the effective magnetization in comparison to the induced anisotropy $H_a<-\Delta M_{eff}$ at temperatures $T\lesssim T_c$.
We found that at $T<T_c$ FMR curves $f_r(H)$ for all studied samples in Tab.~\ref{Tab} obey the modified Kittel dependence: 
\begin{equation}
\begin{aligned}
\left(2\pi f_r/\mu_0\gamma\right)^2= {}& \left(H+H_a+H_s(1-\alpha_s H^2)\right)\times \\
& \times\left(H+H_a+M_{eff}\right)
\end{aligned}
\label{Kit_m}
\end{equation}
where $H_a$ and $M_{eff}$ are constants and are derived at $T>T_c$ (see Tab~\ref{Tab}), $H_s$ is the field of one-dimensional superconducting torque at zero applied field, and the parameter $\alpha_s$ reflects the dependence of the superconducting torque on applied magnetic field.
We confirm that for all studied samples FMR curves at all temperatures $T<T_c$ can be fitted with Eq.~\ref{Kit_m}.
Examples of such fit are shown for the S4 sample in Fig.~\ref{Exp1} with solid lines. 

\begin{figure*}[!ht]
\begin{center}
\includegraphics[width=1\columnwidth]{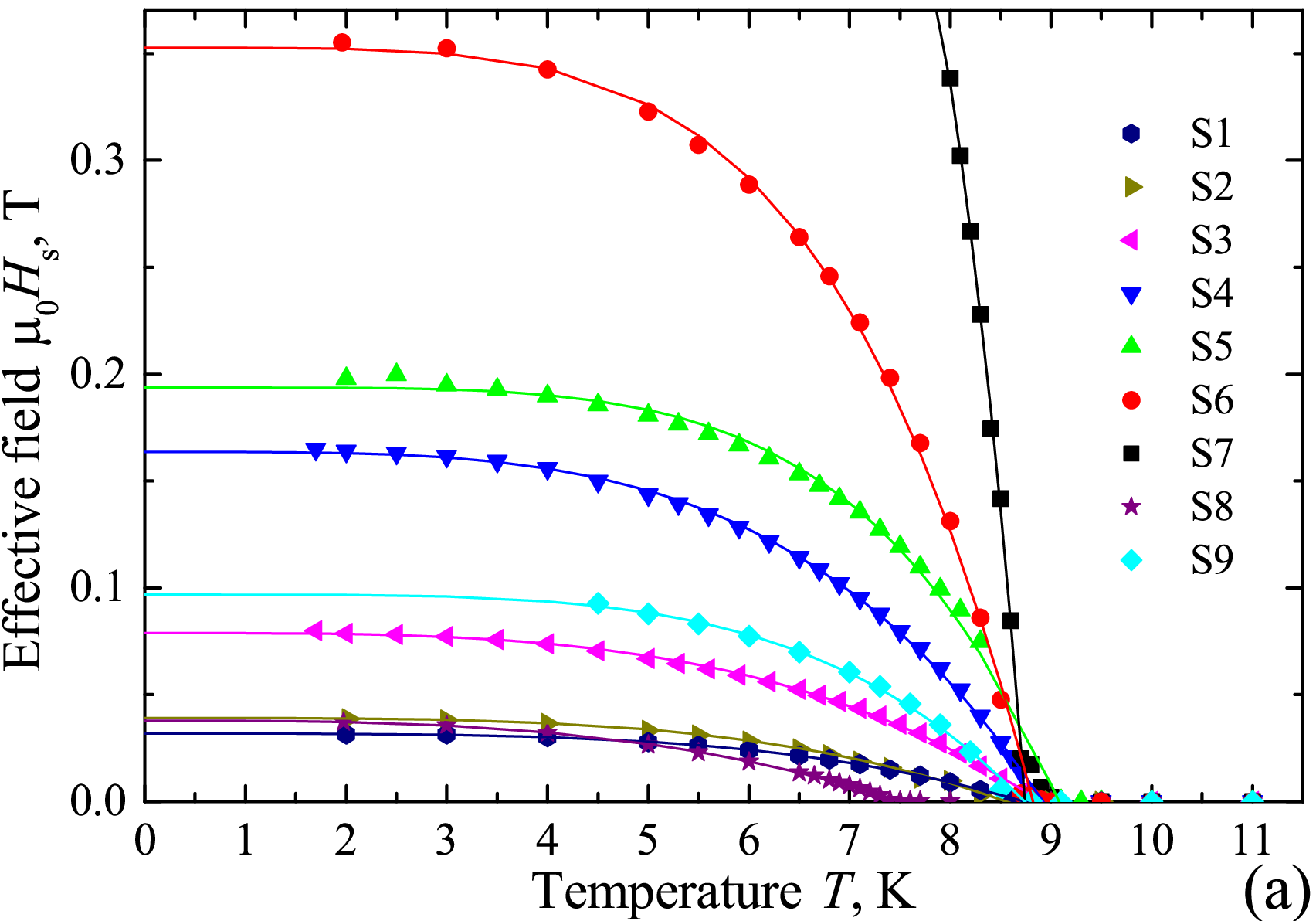}
\includegraphics[width=1\columnwidth]{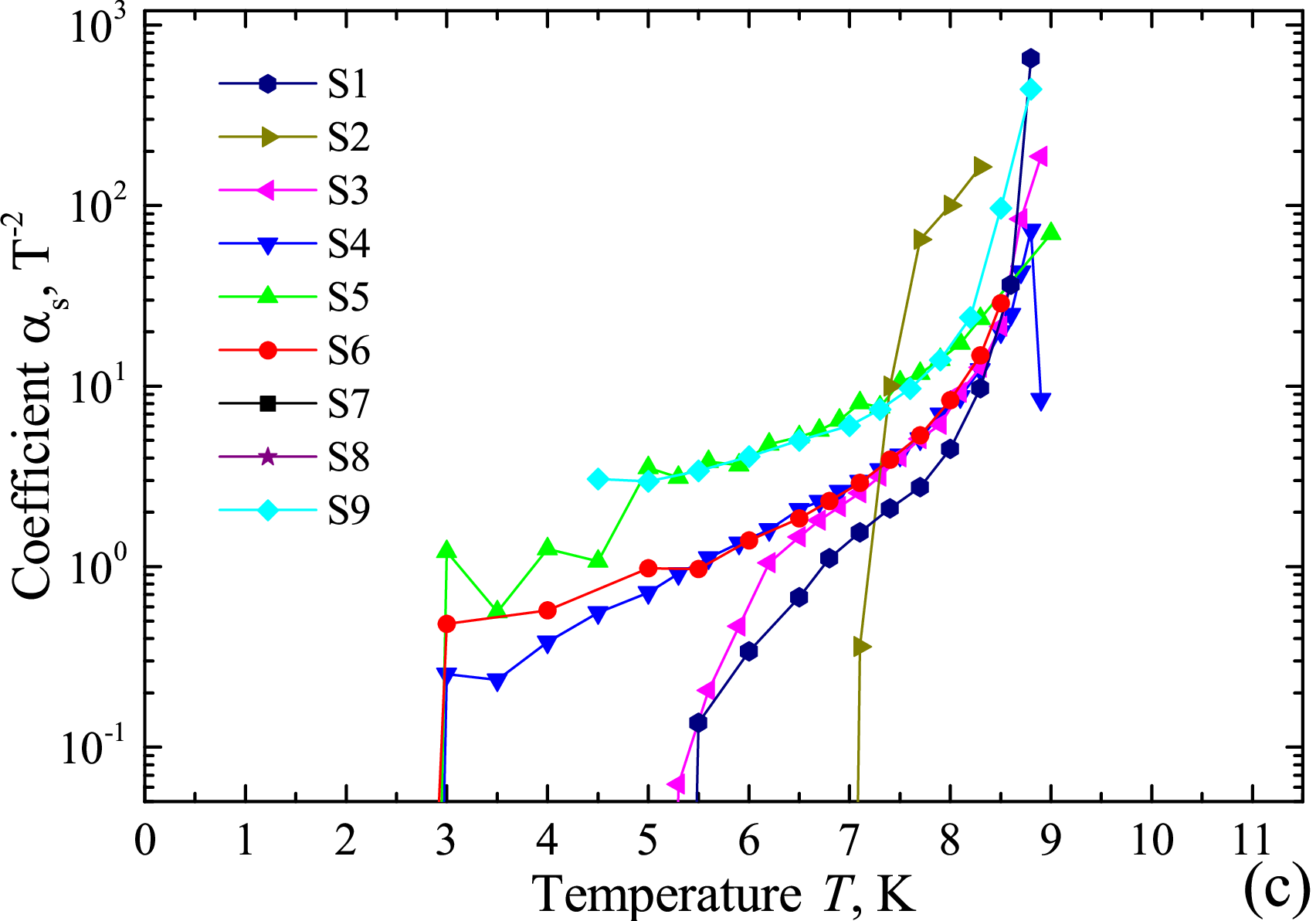}
\includegraphics[width=1\columnwidth]{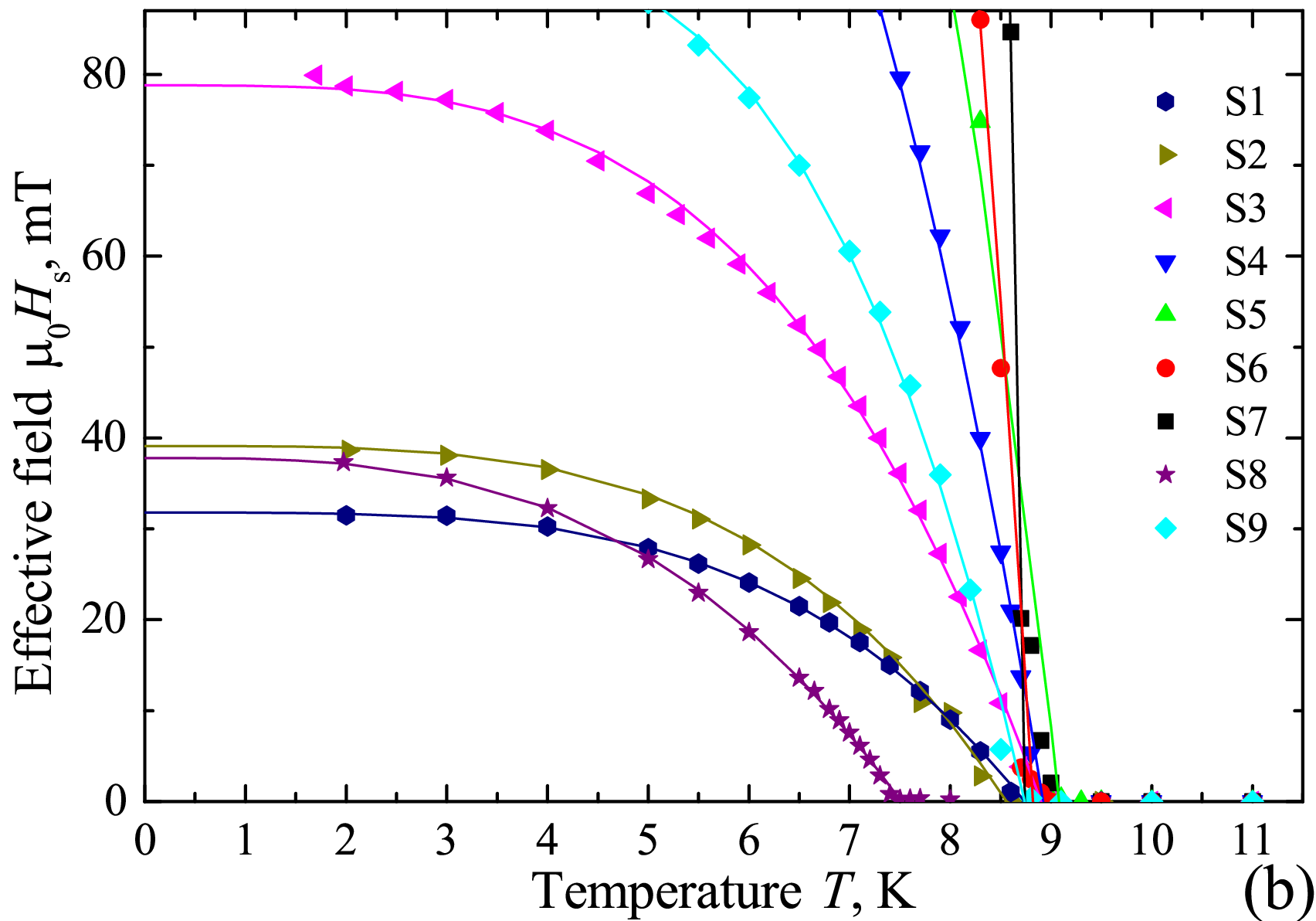}
\includegraphics[width=1\columnwidth]{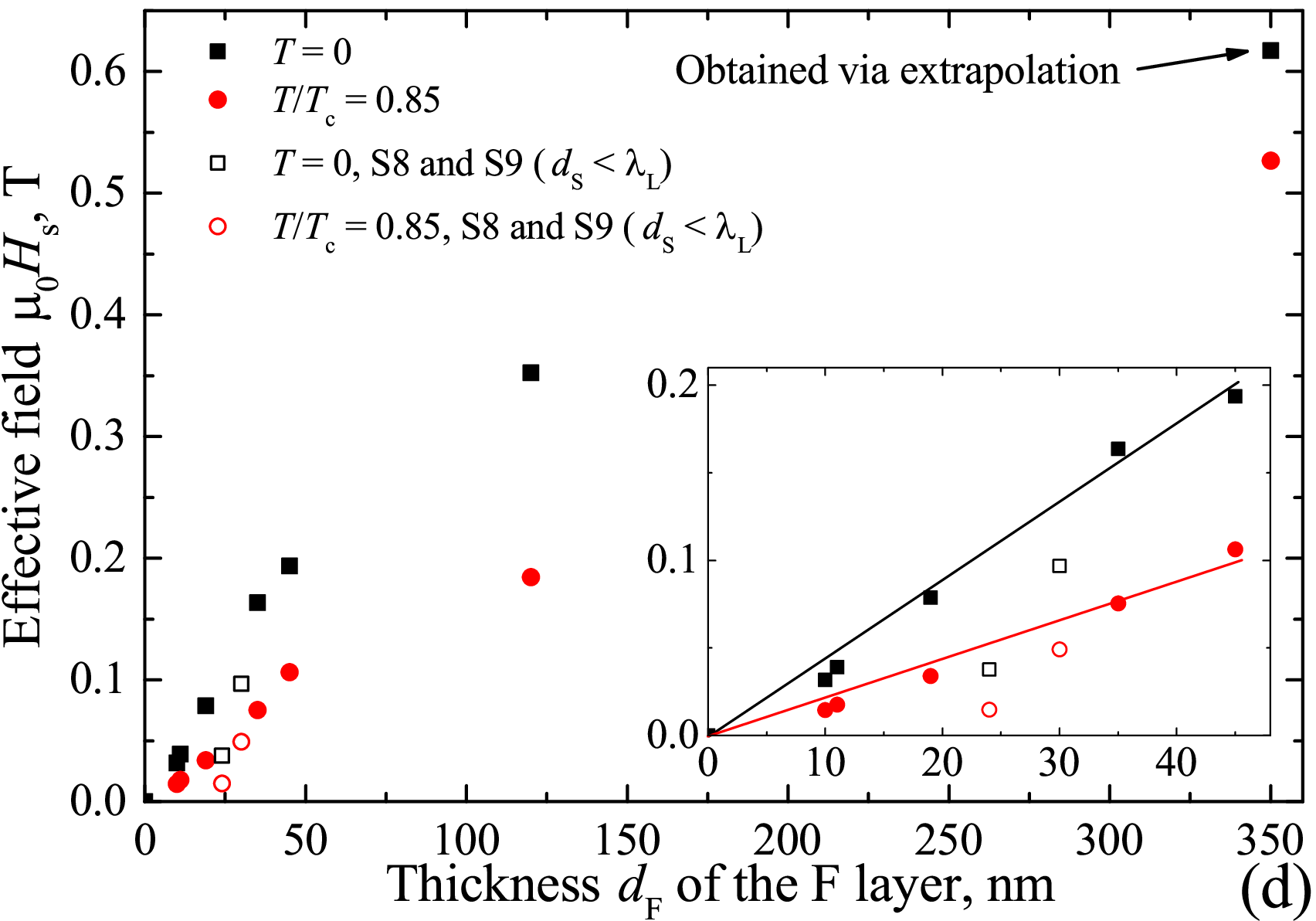}
\caption{
a) Dependencies of the superconducting torque field on temperature $H_s(T)$ at zero external field.
b) Magnification of (a) at $\mu_0 H_s<0.1$~T. 
Solid lines in (a) and (b) show the fit of experimental $H_s(T)$ curves with Eq.~\ref{Hs}.
c) Dependencies of the field-dependence coefficient of the superconducting torque field in Eq.~\ref{Kit_m} on temperature $\alpha_s(T)$.
d) The dependence of the superconducting torque field on the thickness of ferromagnetic layer $H_{s}(d_F)$ at zero field at $T=0$ (black symbols) and at $T/T_c=0.85$ red symbols.
Solid symbols represent data for samples with both superconducting layers $d_s>\lambda_L$.
Open symbols represent data for two samples (S8 and S9) with $d_s<\lambda_L$ at least for one superconducting layer.
See Tab~\ref{Tab} for details. 
The inset in (d) magnifies $H_{s}(d_F)$ for $d_F<45$~nm.
Solid lines in the inset in (d) show linear fit of $H_{s}(d_F)$ at corresponding temperature.
}
\label{Fit1}
\end{center}
\end{figure*}

Analysis of resonance lines $f_r(H)$ with Eqs.~\ref{Kit} and \ref{Kit_m} yields dependencies of the superconducting torque field and of the field-dependence coefficient on temperature, $H_s(T)$ and $\alpha_s(T)$.
This data is provided for all studied samples in Fig.~\ref{Fit1}a-c. 
Notice that for all samples except S8 and S7 the parameter $\alpha_s$ growth exponentially from $\alpha_s\sim 1$~T$^{-2}$ at $T\approx 6$~K up to $\alpha_s\sim 10^2$~T$^{-2}$ at $T\approx 8.5$~K.

Temperature dependence of the torque field $H_s(T)$ can be characterized by fitting it with the following expression \cite{Golovchanskiy_PRAppl_14_024086}:
\begin{equation}
H_s=H_{s0}\left(1-(T/T_c)^p\right)
\label{Hs}
\end{equation}
where $H_{s0}$ is the superconducting spin-torque field at zero field and zero temperature, $T_c$ is the superconducting critical temperature of S-F-S trilayers, and $p$ is a free exponent parameter. 
The fit of $H_{s}(T)$ with Eq.~\ref{Hs} for all samples is shown in Fig.~\ref{Fit1}a-b with solid lines and yields parameters $H_{s0}$ and $p$ provided in Tab~\ref{Tab}.
Notice that for all samples except of the S7, which contains the the thickest F-layer, the exponent $p$ is in the range from about 3 up to 5 with the average value about 4.
Also we notice that due to technical limitations resonance curves of S7 sample with the thickest F-layer could be measured only above to 8~K and the value of $H_s$ below 8~K are obtained via the extrapolation with Eq.~\ref{Hs}. 
This extrapolation yields the natural FMR frequency in the S7 sample $f_r(H=0)=24.1$~GHz at zero temperature in case if the value $p=8.89$ is correct, or $f_r(H=0)=30.0$~GHz for $p=4.9$  (in the latter case $\mu_0 H_{s0}=0.96$~T), which is the maximum value observed among the rest of samples in Tab.~\ref{Tab}.

Figure \ref{Fit1}d shows the dependence of the superconducting torque field on the thickness of ferromagnetic layer $H_{s0}(d_F)$ at zero temperature and at $T/T_c=0.85$.
This is the core result of this work.
Figure \ref{Fit1}d clearly demonstrates the overall logarithmic-like dependence of $H_{s0}$ on thickness $d_F$, with the linear growth of  $H_{s0}$ at low $d_F$ (see the inset in Fig.~\ref{Fit1}d) and retardation of $H_{s0}$ growth at higher $d_F$.
Also, Fig.~\ref{Fit1}d reveals the dependence of the superconducting torque field on the thickness of superconducting layers: in case is the thickness of one of S-layers is reduced $H_{s0}$ is also reduced (open symbols in Fig.~\ref{Fit1}d).
Notice that while in case of the S8 sample this reduction can be partially explained by suppressed superconductivity and smaller $T_c\approx7.5$~K, this is not the case for the S9 sample where $T_c$ is comparable with values for the rest of samples.

\section{Discussion}

The original explanation of the phenomenon in S-F-S trilayers suggested the role of spin-polarized spin-triplet Cooper pairs in formation of the superconductivity-induced anisotropy via the spin-transfer-torque mechanism \cite{Li_ChPL_35_077401}.
Yet, this mechanism requires the frequency of magnetization dynamics of the order of the superconducting gap \cite{Houzet_PRL_101_057009}, can hardly be expected in 350~nm thick junctions, and does not yield the one-dimensional torque on magnetic moment.

The first hint on inductive origin of the phenomenon in S-F-S trilayers was reported in Ref.~\cite{Mironov_APL_119_102601} where the magnetostatic interaction has been considered between the screening currents induced by ferromagnetic stray fields in S-layers and the magnetic moment in the F-layer. 
While only the static case has been considered, it was shown that in case if superconducting currents are allowed to loop around the ferromagnetic layer additional DC demagnetising field is expected with the following dependence on magnetic and structural characteristics of trilayers
\begin{equation}
H_s\propto M_s\frac{d_F}{\lambda_L}\frac{d_F}{L}\ln\frac{2L}{d_F}, 
\label{Mir}
\end{equation}
where $\lambda_L$ is the London penetration depth and $L$ is the length of the structure. 
The theory in Ref.~\cite{Mironov_APL_119_102601} provides an explanation for the one-dimensionality of the demagnetizing field along the $y$-axis in Fig.~\ref{sam} and predicts the growth of the demagnetizing field with increasing thickness of the F-layer.
Yet, according to Eq.~\ref{Mir} this theory predicts a parabolic dependence of $H_s(d_F)$, which is inconsistent with Fig.~\ref{Fit1}d, and some dependence of $H_s$ on the length of the structure, which was not observed experimentally.
In fact, this theory leads to a rather counter-intuitive outcome that the effect in S-F-S structures where superconducting currents are allowed to loop around the ferromagnetic layer should be measurable with conventional magnetization measurements.

The mechanism behind the superconducting torque in S-F-S trilayers was revealed in Ref.~\cite{Silaev}.
The mechanism implies the coupling of the superconducting kinetic inductance with the precessing magnetization at S-F interfaces and formation of macroscopic circulation currents curling around $y$-axis in the opposite phase with precessing magnetization.
The model in Ref.~\cite{Silaev} predicts exactly the same dependence of the resonance frequency on the magnetic field as in Eq.~\ref{Kit_m} in the limit $d_F\ll\lambda$:
\begin{equation}
H_s(1-\alpha_s H^2)=M_s \frac{d_F}{\lambda_L} \tanh \frac{d_S}{\lambda_L}.
\label{Sil}
\end{equation}
In fact, Eq.~\ref{Sil} demonstrates the linear dependence of the superconducting torque on thickness $d_F$ at small $d_F$, which is consistent with Fig.~\ref{Fit1}d.
The reduction of the superconducting torque upon increasing the magnetic field in Eq.~\ref{Sil} is explained by the increase of the penetration depth \cite{Gubin_PRB_72_064503}: $(1-\alpha_s H^2) \propto \nicefrac{1}{\lambda_L(H)} \tanh \nicefrac{d_S}{\lambda_L(H)}$.
The dependence of the superconducting torque on the thickness $d_S$ of S-layers is also captured in Eq.~\ref{Sil} as $\propto \tanh \nicefrac{d_S}{\lambda}$.
Both effects are observed in Fig.~\ref{Fit1}d.
At $d_F\gg\lambda$ the model in Ref.~\cite{Silaev} predicts the saturation of the superconducting torque to a constant value $H_s\approx M_s$, which is also observed in Fig.~\ref{Fit1}d.
Thus, the kinetic inductance origin behind the dramatic increase of the FMR frequency in S-F-S trilayers with electronic interaction at S-F interfaces is confirmed.

As a final remark, it can be expected that superconductivity in S-F-S structures modifies the dispersion of perpendicular standing spin waves (PSSW) \cite{Kittel_PR_100_1295,Seavey_JAP_30_S227,Sparks_PRB_1_3831,Klingler_JPDAP_48_015001,Golovchanskiy_PRMat_6_064406}. 
In the case of closed PSSW boundary conditions \cite{Kittel_PR_100_1295,Seavey_JAP_30_S227,Klingler_JPDAP_48_015001} magnetization precession at both S-F interfaces does not take place and, thus, the superconducting torque is not formed.
In the case of free PSSW boundary conditions \cite{Sparks_PRB_1_3831,Golovchanskiy_PRMat_6_064406} magnetization at S-F interfaces precesses at opposite phases for even modes and precesses in-phase for odd modes, which corresponds to cancellation of the superconducting torque for even modes and its presence for odd modes, respectively.
In the general case when the F-layers is magnetically non-uniform across its thickness \cite{Golovchanskiy_PRMat_6_064406} and the spin boundary conditions are affected by surface anisotropies \cite{Bajorek_JAP_42_4324,Puszkarski_ProgSurfSci_9_191} the effect of the superconducting torue on the dispersion of PSSWs becomes non-trivial.

\section{Conclusion}

Summarising, we report a comprehensive experimental study of magnetization dynamics in S-F-S trilayers. 
We report results of FMR spectroscopy for a large set of samples with varied thickness of both superconducting and ferromagnetic layers in a wide frequency, field, and temperature ranges.
Experimentally we establish an anisotropic one-dimensional action of hybridization-induced torque acting on magnetization dynamics and the dependence of this superconducting torque on the magnetic field.
Experimental results confirm the recently proposed model by M. Silaev \cite{Silaev}, which explains the phenomenon in S-F-S structures as the outcome of the coupling between magnetization dynamics and superconducting kinetic inductance at S-F interfaces.
Our results open wide prospects for application of the superconducting kinetic inductance in magnonics.
In addition, as was suggested in Ref.~\cite{Silaev}, S-F-S systems provide the playground for Anderson-Higgs mass generation of boson quasiparticles in high-energy Standard Model and
in condensed matter systems.

\section{Acknowledgments}

The authors acknowledge Dr. M. Silaev for fruitful discussions, critical reading of the manuscript and useful suggestions.
The research study was financially supported by the Russian Science Foundation (grant N 22-22-00314).

\bibliography{A_Bib_SFS}

\end{document}